\documentclass[a4paper,11pt]{article}

\usepackage[T1]{fontenc} 
\usepackage{float} 

\makeatletter
\gdef\@fpheader{ }
\gdef\@journal{ }
\makeatother
\RequirePackage{amsmath}
\RequirePackage{amssymb}
\RequirePackage{epsfig}
\RequirePackage{graphicx}
\RequirePackage[numbers,sort&compress]{natbib}
\RequirePackage{color}
\RequirePackage[colorlinks=true
,urlcolor=blue
,anchorcolor=blue
,citecolor=blue
,filecolor=blue
,linkcolor=blue
,menucolor=blue
,pagecolor=blue
,linktocpage=true
,pdfproducer=medialab
,pdfa=true
]{hyperref}

\newif\ifnotoc\notocfalse
\newif\ifemailadd\emailaddfalse
\newif\iftoccontinuous\toccontinuousfalse
\makeatletter
\def\@subheader{\@empty}
\def\@keywords{\@empty}
\def\@abstract{\@empty}
\def\@xtum{\@empty}
\def\@dedicated{\@empty}
\def\@arxivnumber{\@empty}
\def\@collaboration{\@empty}
\def\@collaborationImg{\@empty}
\def\@proceeding{\@empty}
\def\@preprint{\@empty}

\newcommand{\subheader}[1]{\gdef\@subheader{#1}}
\newcommand{\keywords}[1]{\if!\@keywords!\gdef\@keywords{#1}\else%
\PackageWarningNoLine{\jname}{Keywords already defined.\MessageBreak Ignoring last definition.}\fi}
\renewcommand{\abstract}[1]{\gdef\@abstract{#1}}
\newcommand{\dedicated}[1]{\gdef\@dedicated{#1}}
\newcommand{\arxivnumber}[1]{\gdef\@arxivnumber{#1}}
\newcommand{\proceeding}[1]{\gdef\@proceeding{#1}}
\newcommand{\xtumfont}[1]{\textsc{#1}}
\newcommand{\correctionref}[3]{\gdef\@xtum{\xtumfont{#1} \href{#2}{#3}}}
\newcommand\jname{JHEP}

\newcommand\preprint[1]{\gdef\@preprint{\hfill #1}}

\newenvironment{proof}[1][Proof]{\noindent\textbf{#1.} }{\ \rule{0.5em}{0.5em}}

\makeatother

\newcommand\note[2][]{%
\if!#1!%
\stepcounter{footnote}\footnotetext{#2}%
\else%
{\renewcommand\thefootnote{#1}%
\footnotetext{#2}}%
\fi}

\makeatletter
\newtoks\auth@toks
\renewcommand{\author}[2][]{%
  \if!#1!%
    \auth@toks=\expandafter{\the\auth@toks#2\ }%
  \else
    \auth@toks=\expandafter{\the\auth@toks#2$^{#1}$\ }%
  \fi
}
\makeatother
\makeatletter
\newtoks\affil@toks\newif\ifaffil\affilfalse
\newcommand{\affiliation}[2][]{%
\affiltrue
  \if!#1!%
    \affil@toks=\expandafter{\the\affil@toks{\item[]#2}}%
  \else
    \affil@toks=\expandafter{\the\affil@toks{\item[$^{#1}$]#2}}%
  \fi
}
\makeatother
\makeatletter
\newtoks\email@toks\newcounter{email@counter}%
\newcommand{\emailAdd}[1]{%
\emailaddtrue%
\ifnum\theemail@counter>0\email@toks=\expandafter{\the\email@toks, \@email{#1}}%
\else\email@toks=\expandafter{\the\email@toks\@email{#1}}%
\fi\stepcounter{email@counter}}
\newcommand{\@email}[1]{\href{mailto:#1}{\tt #1}}
\makeatother

\makeatletter
\newcommand*\collaboration[1]{\gdef\@collaboration{#1}}
\newcommand*\collaborationImg[2][]{\gdef\@collaborationImg{#2}}
\makeatletter
\newcommand\afterLogoSpace{\smallskip}
\newcommand\afterSubheaderSpace{\vskip3pt plus 2pt minus 1pt}
\newcommand\afterProceedingsSpace{\vskip21pt plus0.4fil minus15pt}
\newcommand\afterTitleSpace{\vskip23pt plus0.06fil minus13pt}
\newcommand\afterRuleSpace{\vskip23pt plus0.06fil minus13pt}
\newcommand\afterCollaborationSpace{\vskip3pt plus 2pt minus 1pt}
\newcommand\afterCollaborationImgSpace{\vskip3pt plus 2pt minus 1pt}
\newcommand\afterAuthorSpace{\vskip5pt plus4pt minus4pt}
\newcommand\afterAffiliationSpace{\vskip3pt plus3pt}
\newcommand\afterEmailSpace{\vskip16pt plus9pt minus10pt\filbreak}
\newcommand\afterXtumSpace{\par\bigskip}
\newcommand\afterAbstractSpace{\vskip16pt plus9pt minus13pt}
\newcommand\afterKeywordsSpace{\vskip16pt plus9pt minus13pt}
\newcommand\afterArxivSpace{\vskip3pt plus0.01fil minus10pt}
\newcommand\afterDedicatedSpace{\vskip0pt plus0.01fil}
\newcommand\afterTocSpace{\bigskip\medskip}
\newcommand\afterTocRuleSpace{\bigskip\bigskip}
\newlength{\affiliationsSep}
\newcommand\beforetochook{\pagestyle{myplain}\pagenumbering{roman}}

\DeclareFixedFont\trfont{OT1}{phv}{b}{sc}{11}

\renewcommand\maketitle{
\pagestyle{empty}
\thispagestyle{titlepage}
\setcounter{page}{0}
\noindent{\small\scshape\@fpheader}\@preprint\par

\afterLogoSpace
\if!\@subheader!\else\noindent{\trfont{\@subheader}}\fi
\afterSubheaderSpace
\if!\@proceeding!\else\noindent{\sc\@proceeding}\fi
\afterProceedingsSpace
{\LARGE\flushleft\sffamily\bfseries\@title\par}
\afterTitleSpace
\hrule height 1.5\p@%
\afterRuleSpace
\if!\@collaboration!\else
{\Large\bfseries\sffamily\raggedright\@collaboration}\par
\afterCollaborationSpace
\fi
\if!\@collaborationImg!\else
{\normalsize\bfseries\sffamily\raggedright\@collaborationImg}\par
\afterCollaborationImgSpace
\fi
{\bfseries\raggedright\sffamily\the\auth@toks\par}
\afterAuthorSpace
\ifaffil\begin{list}{}{%
\setlength{\leftmargin}{0.28cm}%
\setlength{\labelsep}{0pt}%
\setlength{\itemsep}{\affiliationsSep}%
\setlength{\topsep}{-\parskip}}
\itshape\small%
\the\affil@toks
\end{list}\fi
\afterAffiliationSpace
\ifemailadd 
\noindent\hspace{0.28cm}\begin{minipage}[l]{.9\textwidth}
\begin{flushleft}
\textit{E-mail:} \the\email@toks
\end{flushleft}
\end{minipage}
\else 
\PackageWarningNoLine{\jname}{E-mails are missing.\MessageBreak Plese use \protect\emailAdd\space macro to provide e-mails.}
\fi
\afterEmailSpace
\if!\@xtum!\else\noindent{\@xtum}\afterXtumSpace\fi
\if!\@abstract!\else\noindent{\renewcommand\baselinestretch{.9}\textsc{Abstract:}}\ \@abstract\afterAbstractSpace\fi
\if!\@keywords!\else\noindent{\textsc{Keywords:}} \@keywords\afterKeywordsSpace\fi
\if!\@arxivnumber!\else\noindent{\textsc{ArXiv ePrint:}} \href{http://arxiv.org/abs/\@arxivnumber}{\@arxivnumber}\afterArxivSpace\fi
\if!\@dedicated!\else\vbox{\small\it\raggedleft\@dedicated}\afterDedicatedSpace\fi
\ifnotoc\else
\iftoccontinuous\else\newpage\fi
\beforetochook\hrule
\tableofcontents
\afterTocSpace
\hrule
\afterTocRuleSpace
\fi
\setcounter{footnote}{0}
\pagestyle{myplain}\pagenumbering{arabic}
} 

\renewcommand{\baselinestretch}{1.1}\normalsize
\divide\@tempdima\baselineskip
\@tempcnta=\@tempdima
\skip\@mpfootins = \skip\footins
\renewcommand{\@dotsep}{10000}

\newcommand\ps@myplain{
\pagenumbering{arabic}
\renewcommand\@oddfoot{\hfill-- \thepage\ --\hfill}
\renewcommand\@oddhead{}}
\let\ps@plain=\ps@myplain

\newcommand\ps@titlepage{\renewcommand\@oddfoot{}\renewcommand\@oddhead{}}

\numberwithin{equation}{section}

\renewcommand\section{\@startsection{section}{1}{\z@}%
                                   {-3.5ex \@plus -1.3ex \@minus -.7ex}%
                                   {2.3ex \@plus.4ex \@minus .4ex}%
                                   {\normalfont\large\bfseries}}
\renewcommand\subsection{\@startsection{subsection}{2}{\z@}%
                                   {-2.3ex\@plus -1ex \@minus -.5ex}%
                                   {1.2ex \@plus .3ex \@minus .3ex}%
                                   {\normalfont\normalsize\bfseries}}
\renewcommand\subsubsection{\@startsection{subsubsection}{3}{\z@}%
                                   {-2.3ex\@plus -1ex \@minus -.5ex}%
                                   {1ex \@plus .2ex \@minus .2ex}%
                                   {\normalfont\normalsize\bfseries}}
\renewcommand\paragraph{\@startsection{paragraph}{4}{\z@}%
                                   {1.75ex \@plus1ex \@minus.2ex}%
                                   {-1em}%
                                   {\normalfont\normalsize\bfseries}}
\renewcommand\subparagraph{\@startsection{subparagraph}{5}{\parindent}%
                                   {1.75ex \@plus1ex \@minus .2ex}%
                                   {-1em}%
                                   {\normalfont\normalsize\bfseries}}

\def\fnum@figure{\textbf{\figurename\nobreakspace\thefigure}}
\def\fnum@table{\textbf{\tablename\nobreakspace\thetable}}

\long\def\@makecaption#1#2{%
  \vskip\abovecaptionskip
  \sbox\@tempboxa{\small #1. #2}%
  \ifdim \wd\@tempboxa >\hsize
    \small #1. #2\par
  \else
    \global \@minipagefalse
    \hb@xt@\hsize{\hfil\box\@tempboxa\hfil}%
  \fi
  \vskip\belowcaptionskip}

\renewenvironment{thebibliography}[1]{%
\begin{oldthebibliography}{#1}%
\small%
\raggedright%
\setlength{\itemsep}{5pt plus 0.2ex minus 0.05ex}%
}%
{%
\end{oldthebibliography}%
}

\begin{document}


\title{\boldmath Converting Lattices into Networks: 
The Heisenberg Model and Its 
Generalizations with Long-Range Interactions}


\author[a,d]{Chi-Chun Zhou,}\note{zhouchichun@dali.edu.cn}
\author[b]{Yao Shen,}\note{shenyaophysics@hotmail.com. Corresponding author.}
\author[c,d]{Yu-Zhu Chen,}\note{chenyuzhu@tju.edu.cn}
\author[d]{and Wu-sheng Dai}\note{daiwusheng@tju.edu.cn}
\

\affiliation[a]{School of Engineering, Dali University, Dali, Yunnan 671003, PR China}
\affiliation[b]{School of Criminal Investigation, People's Public Security University of China, Beijing 100038, PR China  }
\affiliation[c]{Theoretical Physics Division, Chern Institute of Mathematics, Nankai University, Tianjin, 300071, PR China}
\affiliation[d]{Department of Physics, Tianjin University, Tianjin 300350, PR China}










\abstract{In this paper, we convert the lattice configurations
into networks with different modes of links and 
consider models on networks with arbitrary numbers of interacting particle-pairs. 
We solve the Heisenberg model
by revealing the relation between the Casimir operator 
of the unitary group and the
conjugacy-class operator of the permutation group.
We generalize the Heisenberg model by this relation and
give a series of exactly solvable models.
Moreover, by numerically calculating the eigenvalue of Heisenberg models and 
random walks on network with different numbers of links,
we show that a system on lattice configurations with interactions 
between more particle-pairs 
have higher degeneracy of eigenstates. 
The highest degeneracy of eigenstates
of a lattice model is discussed.}

\maketitle
\flushbottom


\section{Introduction}
Lattice models are important because they 
can be not only treated theoretically, 
for example the Heisenberg models 
can be solved exactly at lower dimensions 
\cite{baxter2016exactly,reichl2009modern,pathria2011statistical}, 
but also implemented experimentally, such as constraining 
ultracold atoms in optical lattices
\cite{galitski2013spin,goldman2014light}. 
Among many lattice models, 
the Heisenberg model which describes 
the magnetism of a system \cite{baxter2016exactly,reichl2009modern,pathria2011statistical} 
is of significance. 
To solve the Heisenberg model, methods such
as the Bethe Ansatz and the Jordan Wigner transformations 
are proposed \cite{baxter2016exactly,crampe2011coordinate}.
The solution of the Heisenberg model at different dimensions
\cite{arnesen2001natural,sandvik1997finite,bougourzi1996exact} or under
certain boundary conditions \cite{belliard2013heisenberg,lauchli2006quadrupolar,sindzingre2010phase} 
is discussed.

In this paper, we solve the Heisenberg model by using a group theory method. 
In the group theory method suggested in the present paper, we reveal 
the relation between the Casimir operator of the unitary group and the
conjugacy-class operator of the permutation group. 
We propose a generalization of the Heisenberg model  
which can be solved exactly. 

\begin{figure}[H]
\centering
\includegraphics[width=1.0\textwidth]{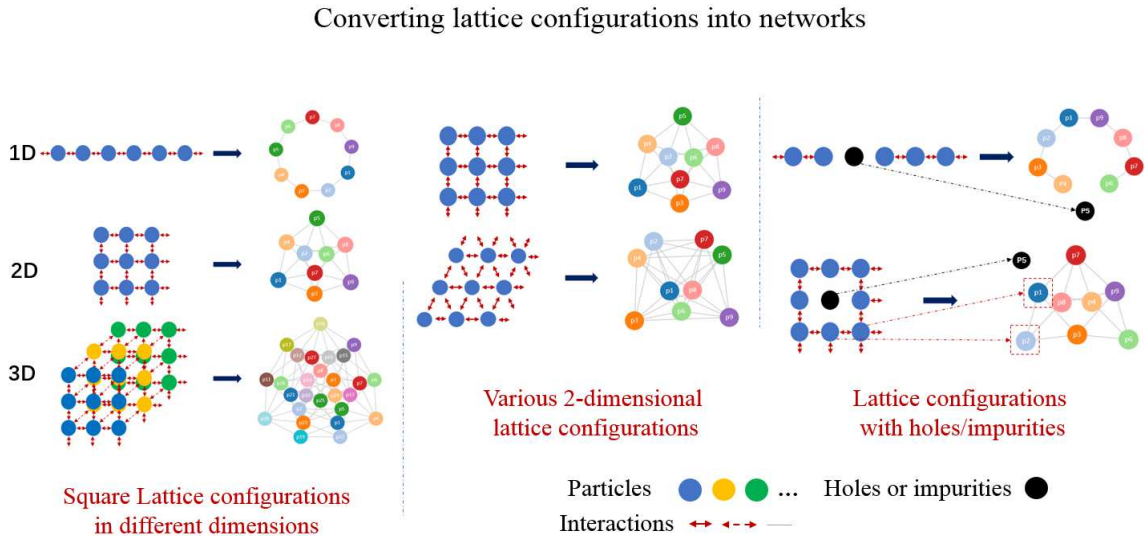}
\caption{Converting lattice configurations into networks.
By holes or impurities, we mean that they can not interact with 
other particles. The periodic boundary condition is applied.}
\label{Topological_equivalence}
\end{figure}

\textit{(1) Converting lattice configurations into 
networks.} We consider a model on lattice configurations 
with arbitrary numbers of interacting particle-pairs by 
converting the lattice configuration into networks.
Usually, lattice models with certain number 
of neighbours or interacting particle-pairs are taken into account.
For example, $3$, $4$, $6$, etc.
It is because only countable lattice configurations exist under the 
constraint of symmetries \cite{braess2007finite,zienkiewicz2000finite}. 
For example, there are $17$ kinds 
of two-dimensional lattice models 
and there are $230$ kinds of three-dimensional 
lattice models. In this paper, we map the lattice configuration to networks 
with different modes of links, as shown in Fig. (\ref{Topological_equivalence}).
We show that
the number of nearest particles is characterized by the 
number of links in the network.
The advantage of doing this is that 
we are free to consider a systems 
with arbitrary number of interacting particle-pairs which in lattice configurations is 
restricted by symmetries. 
By studying the relation between degeneracy of eigenstates and 
the number of the interacting particle-pairs, 
we show that models on lattice configuration with more interacting particle-pairs 
have higher degeneracy of eigenstates.
Thus, lattice models with interactions between all particle-pairs 
should have a highest degeneracy of eigenstates.

\textit{(2) The Long-Range Interactions.}
The lattice model with long-range interaction between 
inter-particles is considered. 
Usually, short-range interactions are 
considered in lattice models, as a result,
only the interaction between nearest, 
second nearest, third nearest, etc, 
particle-pairs need to be considered. 
However, if long-range interactions, such as the inverse square potential 
and the harmonic oscillator potential, exist,
interactions between more particle-pairs need to be considered.
For example, one-dimensional models with long-range 
interactions is considered in \cite{feynman1998statistical}.
Ising models on the hypercube lattices with long-range interaction is considered in \cite{litinskii2020eigenvalues}.
In this paper, the model on lattice 
configuration with interaction
between all particle-pairs is with long-range interactions.

This paper is organized as follows: 
In Sec. 2, we solve 
the Heisenberg model on lattice configurations with  
interactions between all particle-pairs by revealing the 
relation between the Casimir operator 
of the unitary group and the
conjugate class operator of the permutation group. 
In Sec. 3, we give a series of exactly solvable models by generalizing the Heisenberg model from
this relation. 
In Sec. 4, we investigate the relation between degeneracy of eigenstates 
and the number of links in a network by numerically calculating 
the eigenvalues of Heisenberg models and random walks on a network.
We discuss the highest degeneracy of eigenstates for a interacting many-body system.
Conclusions and discussions are given in Sec. 5. 

\section{The Heisenberg model on lattice configurations with interactions between all particle-pairs}
In this section, we consider the Heisenberg model 
on lattice configurations with 
interactions between all particle-pairs. 

\begin{figure}[H]
\centering
\includegraphics[width=1.0\textwidth]{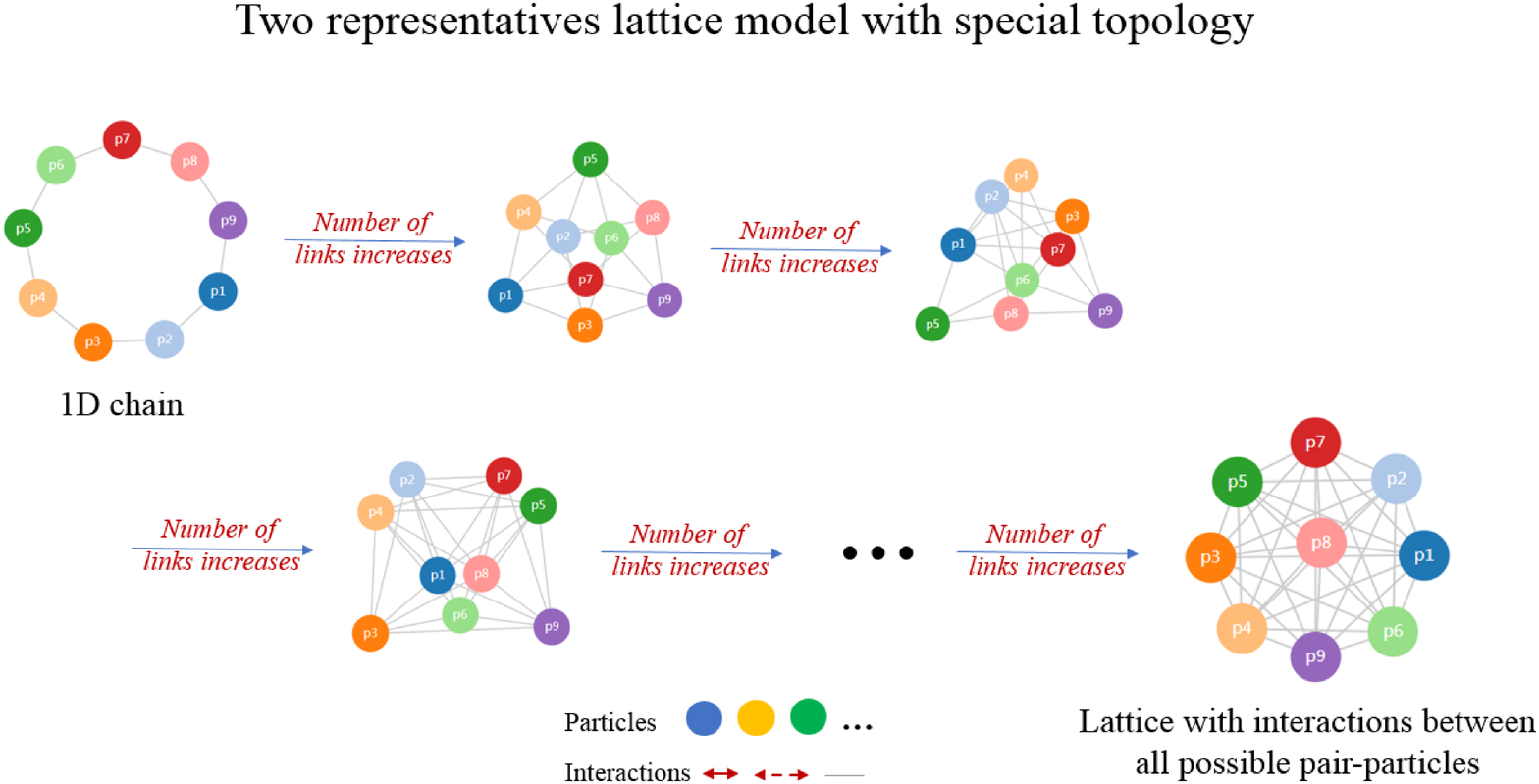}
\caption{The lattice configuration with interactions between all
particle-pairs and the one-dimensional lattice configuration are two 
special lattice configurations:
one considers the largest number of links and the other considers the smallest number of links.}
\label{two_topo}
\end{figure}

\textit{A special lattice configuration: lattice configurations with 
interactions between all particle-pairs.} As shown in Fig. 
(\ref{Topological_equivalence}), for various kinds of lattice configurations 
the difference varies
in the mode of links between particles other than the position of particles. 
The network with links between all particle-pairs
considers the largest number of particle-pairs interactions, as shown in Fig. 
(\ref{two_topo}). In the following, 
we will make no distinguish between lattice configurations and networks.

\textit{The Hamiltionian of the Heisenberg model with 
interactions between all particle-pairs.} 
Here, we show that the Hamiltonian of the Heisenberg model on the lattice configuration
with interactions between all particle-pairs
can be expressed in terms of
the Casimir operators of $U\left(  2\right)  $ group. 
We give the eigenvalue spectrum of the system.
For the sake of clarity, in this section, we directly give the result.
The detail of calculations will be given in Sec. 3. 
For the Heisenberg model on lattice configurations with interactions between
all particle-pairs, the Hamiltonian reads%
\begin{equation}
H=\sum_{\left(  i,j\right)  }\frac{1}{2}\mathbf{S}_{i}\cdot\mathbf{S}_{j},\label{60}%
\end{equation}
where $\mathbf{S}_{i}$ is the spin of $ith$ particle and $\sum_{\left(
i,j\right)  }$ runs all pairs of particles. With
simple transformations, the Hamiltonian, Eq. (\ref{60}), can be written as%
\begin{equation}
H=\sum_{\left(  i,j\right)  }\tau_{ij}-\sum_{\left(  i,j\right)  }%
\mathbf{e}\cdot\mathbf{e},\label{600}%
\end{equation}
where $\tau_{ij}$ if the action of permutation of particles with index $i$ and $j$.
and $\mathbf{e}$ is the identity matrix. 
By using the relation between the Casimir operator of unitary group 
and the conjugacy-class operator of permutation group, which, for the sake of clarity, 
will be provided in Sec. 3.1,
one can rewrite Eq. (\ref{600}) in the form%
\begin{equation}
H\mathbf{=}P_{\left(  1^{N-2},2\right)}=\frac{1}{2}C_{2}-C_{1},\label{68}%
\end{equation}
where $C_{2}$ and $C_{1}$ are the Casimir operator of the unitary group 
$U\left(  2\right)  $ with order $2$ and $1$, respectively.
$P_{\left(  1^{N-2},2\right)}$ is the conjugacy-class operator of the symmetrical group.

\textit{The eigenvalue spectrum.} 
The Hamiltionain, Eq. (\ref{68}), shows that the eigenvalue of the Hamiltionain can be obtained once 
the eigenvalues of the Casimir operator of the unitary group $U\left(  2\right)  $ are given.
For a system consisting of $N$ particles, 
the irreducible representation
of $U\left(  2\right)  $ is indexed by a single parameter $s$, 
an integer ranges from $0$ to the largest integer smaller than $N/2$
\cite{Iachello2006Lie}. By using the eigenvalue of the Casimir operator of $U\left(  2\right)  $
in a given representation \cite{Iachello2006Lie}, 
we give the eigenvalue of the Hamiltionain, Eq.
(\ref{68}):%
\begin{equation}
E_{s}=2s^{2}-2Ns-2s+\frac{1}{2}N^{2}-\frac{1}{2}N.
\end{equation}
The degeneracy of energy $E_{s}$ is $\omega_{s}$%
\begin{equation}
\omega_{s}=\frac{\left(  1+N-2s\right)  ^{2}N!}{\left(  1+N-s\right)  !s!}.
\end{equation}

\section{A generalization of the Heisenberg model}

In this section, we construct a series of exactly 
solvable models. These models
are the generalization of the Heisenberg model 
using the relation between
the Casimir operator of the unitary group
and the conjugacy-class operator of permutation group. 
This method provides us an easier way to give the spectrum of the many-body 
spin system. As shown in Sec. 2, setting $m=2$ recovers 
the Heisenberg model. 
For the generalized model,
bosonic and fermionic realizations are discussed below.

\subsection{The Casimir operator of the unitary group and the 
conjugacy-class operator of the permutation group: a brief review}

\textit{The Casimir operator of the unitary group }$U\left(  m\right)
$\textit{.} The unitary group $U\left(  m\right)  $ has $m$
linear-independent Casimir operators. The Casimir
operator of order $l$, denoted by $C_{l}$, is \cite{Iachello2006Lie}
\begin{equation}
C_{l}=\sum_{k_{1},k_{2},...,k_{l-1}}E_{k_{1}k_{2}}E_{k_{2}k_{3}}%
...E_{k_{l-2}k_{l-1}}E_{k_{l-1}k1}\label{62}%
\end{equation}
where $E_{kl}$ is the generator of $U\left(  m\right)  $
.

The irreducible
representation of $U\left(  m\right)  $ is indexed by $m$ nonnegative
integers denoted by $\left(  a\right)  =\left(  a_{1},a_{2},\ldots
,a_{m}\right)  $ with $a_{1}\geqslant a_{2}\geqslant...\geqslant
a_{m}\geqslant0$ \cite{Iachello2006Lie}. 
The eigenvalue of the Casimir operator $C_{l}$,
denoted by $\left\langle C_{1}\right\rangle$, under the
representation indexed by $\left(  a\right)  $ is \cite{Iachello2006Lie}
\begin{align}
\left\langle C_{1}\right\rangle  & =S_{1}\label{c1}\\
\left\langle C_{2}\right\rangle  & =S_{2}-\left(  m-1\right)  S_{1}%
\label{c2}\\
\left\langle C_{3}\right\rangle  & =S_{3}-\left(  m-\frac{3}{2}\right)
S_{2}-\frac{1}{2}S_{1}^{2}-\left(  m-1\right)  S_{1}\\
& \ldots,\label{c4}%
\end{align}
where
\begin{equation}
S_{k}=\sum_{i=1}^{m}\left[  \left(  a_{i}+m-i\right)  ^{k}-\left(  m-i\right)
^{k}\right]  .
\end{equation}

\textit{The conjugacy-class operator of the permutation group }$S_{N}%
$\textit{.} The sum of all the group elements which belong to the same conjugacy class
gives a conjugacy-class operator \cite{chen2002group}. The conjugacy-class
operator commutes with all the group elements \cite{chen2002group}. For
the permutation group of order $N$, each integer partition of $N$ (a
representation of $N$ in terms of the sum of other integers) gives a corresponding
conjugacy-class operator  \cite{chen2002group}. In this paper, we focus on the conjugacy-class
operator corresponding to the integer partition $\left(  \lambda\right)
=\left(  1^{N-2},2\right)  $ with superscript $N-2$ standing for $1$ appearing
$N-2$ times. By definition, the conjugacy-class operator reads%
\begin{equation}
P_{\left(  1^{N-2},2\right)  }=\sum_{1=i<j}^{N}\tau_{ij},\label{61}%
\end{equation}
where $\tau_{ij}$ is the exchange of the $ith$ and the $jth$ particle.

\subsection{A relation between the Casimir operator of the unitary group and the
conjugate class operator of the permutation group}

In this section, we introduce a relation between the Casimir 
operator of the unitary group and the conjugacy-class operator of the permutation group. 
This relation makes it easier to calculate the eigenvalue spectrum of the Heisenberg model .

\textit{The relation between }$C_{l}$\textit{ and
}$P_{\left(  1^{N-2},2\right)  }$. The conjugacy-class operator $P_{\left(  1^{N-2},2\right)  }$ of the permutation group satisfies%
\begin{equation}
P_{\left(  1^{N-1},2\right)  }=\frac{1}{2}C_{2}-\frac{m}{2}C_{1},\label{67}%
\end{equation}
where $C_{1}$ and $C_{2}$ are the Casimir operators of order $1$ and $2$ of the unitary group.

\begin{proof}
Let $a_{k}^{\dagger i}$ with superscript ranging from $1$ to $N$ and
subscript ranging from $1$ to $m$ represents creating the $ith$
particle in the state $k$. $a_{k}^{i}$ is the annihilation operator, which represents
the conjugate of $a_{l}^{\dagger j}$. The commutation relation between
$a_{k}^{\dagger i}$ and $a_{l}^{j}$ is%
\begin{align}
\left[  a_{k}^{i},a_{l}^{\dagger j}\right]   &  =\delta_{ij}\delta_{kl},\\
\left[  a_{k}^{i},a_{l}^{j}\right]   &  =\left[  a_{k}^{\dagger i}%
,a_{l}^{\dagger j}\right]  =0.
\end{align}

The generator $\tau_{ij}$ of $S_{N}$ can be expressed in terms of $a_{l}%
^{j}$ and $a_{l}^{\dagger j}$ \cite{chen2002group}:
\begin{equation}
\tau_{ij}=\sum_{k,l=1}^{m}a_{k}^{\dagger i}a_{l}^{\dagger j}a_{l}^{i}a_{k}%
^{j}.\label{63}%
\end{equation}
The generator of $U\left(  m\right)  $ can also be expressed as:
\begin{equation}
E_{kl}=\sum_{i=1}^{N}a_{k}^{\dagger i}a_{l}^{i}.\label{64}%
\end{equation}
Substituting Eq. (\ref{63}) with $\tau_{ij}$ into Eq. (\ref{61}) gives the conjugacy-class
operator:
\begin{equation}
P_{\left(  1^{N-2},2\right)  }=\sum_{1=i<j}^{N}\sum_{k,l=1}^{m}a_{k}^{\dagger
i}a_{l}^{\dagger j}a_{l}^{i}a_{k}^{j}.\label{642}%
\end{equation}
Substituting Eq. (\ref{64}) with $E_{kl}$ into Eq. (\ref{62}) gives the Casimir operator:%
\begin{align}
C_{2} &  =\sum_{i,j}^{N}\sum_{k,l=1}^{m}a_{k}^{\dagger i}a_{l}^{\dagger
j}a_{l}^{i}a_{k}^{j}C,\label{651}\\
C_{1} &  =\sum_{i=1}^{N}\sum_{l=1}^{m}a_{l}^{\dagger i}a_{l}^{i}.\label{652}%
\end{align}
So the operator relation between $P_{\left(  1^{N-2},2\right)  }$ and $C_{l}$ becomes
\begin{equation}
\frac{1}{2}C_{2}-\frac{m}{2}C_{1}=P_{\left(  1^{N-1},2\right)  }+\frac{1}%
{2}\sum_{k,l=1}^{m}\sum_{i=1}^{N}a_{k}^{\dagger i}a_{l}^{\dagger i}a_{l}%
^{i}a_{k}^{i},\label{6523}%
\end{equation}
where Eq. (\ref{642}) is used. The last term of Eq. (\ref{6523}) can be
written as
\begin{equation}
\sum_{k,l=1}^{m}\sum_{i=1}^{N}a_{k}^{\dagger i}a_{l}^{\dagger i}a_{l}^{i}%
a_{k}^{i}=\sum_{i=1}^{N}\left[  \sum_{l=1}^{m}n_{l}^{i}-\left(  \sum_{l=1}%
^{m}n_{l}^{i}\right)  \left(  \sum_{l=1}^{m}n_{l}^{i}\right)  \right]  ,
\end{equation}
by introducing $N_{i}\equiv\sum_{l=1}^{m}n_{l}^{i}$ with $n_{l}^{i}$ equaling to
$a_{l}^{\dagger i}a_{l}^{i}$. $N_{i}$ represents the number of $ith$ particle,
that is $1$. Therefore, the equation%
\begin{equation}
\sum_{k,l=1}^{m}\sum_{i=1}^{N}a_{k}^{\dagger i}a_{l}^{\dagger i}a_{l}^{i}%
a_{k}^{i}=0\label{653}%
\end{equation}
holds.  Eq. (\ref{67}) is proved.
\end{proof}

\subsection{The generalization}
In this section, by using the relation between the Casimir 
operator of the unitary group and the conjugacy-class operator of the permutation group, 
Eq. (\ref{67}),
we construct a series of exactly solvable models which are generalizations of the Heisenberg model.
They are models on lattice configurations with interactions between all particle-pairs.

\textit{The Hamiltionian.} The Hamiltonian of the model is
\begin{equation}
H=\frac{1}{2}C_{2}-\frac{m}{2}C_{1} \label{haha}%
\end{equation}
with $C_{1}$ and $C_{2}$ the Casimir operators of $U\left(  m\right)  $. 
Such systems consist of $N$ $m$-states particles.
The Hamiltonian $H$ in Eq. (\ref{haha}) 
exchanges all particle-pairs in the system, that is
\begin{equation}
H\left(  v_{1}\otimes v_{2}\otimes v_{3}\otimes...\otimes v_{N}\right)
=v_{2}\otimes v_{1}\otimes v_{3}\otimes...\otimes v_{N}+v_{1}\otimes
v_{3}\otimes v_{2}\otimes...\otimes v_{N}+\ldots.
\end{equation}
Therefore, the generalized models are on 
lattice configurations with interactions between all particle-pairs.

\textit{The eigenvalue spectrum.} 
Eq. (\ref{haha}) shows that, the eigenvalue of the Hamiltionain can be obtained provided 
the eigenvalues of the Casimir operator of the unitary group $U\left(  m\right)  $ are given.
The eigenvalues of the Casimir operator of the unitary group $U\left(  m\right)$ 
are indexed by $m$
nonnegative integers denoted by $\left(  a\right)  =\left(  a_{1}%
,a_{2},...,a_{m}\right)  $ with $a_{1}\geqslant a_{2}\geqslant...\geqslant
a_{m}\geqslant0$ \cite{Iachello2006Lie}. 
The summation of $\left(  a\right)  $ is $N$, i.e.,
$\sum_{i=1}^{m}a_{i}=N$. By using the eigenvalue of the Casimir operator given
in Eqs. (\ref{c1})-(\ref{c2}), we give the eigenvalue of the Hamiltionain, Eq.
(\ref{haha}),
\begin{equation}
E_{\left(  a\right)  }=\frac{1}{2}\sum_{i=1}^{m}a_{i}^{2}-\sum_{i=1}^{m}%
a_{i}i+\frac{1}{2}N.\label{610}%
\end{equation}
The degeneracy of $E_{\left(  a\right)  }$ is%
\begin{equation}
\omega_{\left(  a\right)  }=N!%
{\displaystyle\prod\limits_{i<j}^{m}}
\frac{\left(  a_{i}-a_{j}-i+j\right)  }{\left(  j-i\right)  }\frac{%
{\displaystyle\prod\limits_{i^{\prime}<j^{\prime}}^{l}}
\left(  a_{i^{\prime}}-a_{j^{\prime}}-i^{\prime}+j^{\prime}\right)  }{%
{\displaystyle\prod\limits_{i^{\prime\prime}=1}^{l}}
\left(  l+a_{i^{\prime\prime}}-i^{\prime\prime}\right)  !},\label{611}%
\end{equation}
where $l$ is the number of positive integers in $\left(  a\right)  $.

\textit{A bosonic realization: an example. }
By expressing the generators of $U\left(  m\right)  $
in terms of the Boson operators $a_{i}^{\dagger}$ and $a_{j}$ with commutation
relation%
\begin{align}
\left[  a_{i},a_{j}^{\dagger}\right]   &  =\delta_{ij},\\
\left[  a_{k},a_{j}\right]   &  =\left[  a_{k}^{\dagger},a_{l}^{\dagger
}\right]  =0,
\end{align}
the Hamiltonian of the system, Eq. (\ref{haha}), can be written as
\begin{equation}
H=\frac{1}{2}\sum_{i,j=1}^{m}a_{i}^{\dagger}a_{j}a_{j}^{\dagger}a_{i}-\frac
{m}{2}\sum_{i=1}^{m}a_{i}^{\dagger}a_{I}.%
\end{equation}
By using Eqs. (\ref{610}) and (\ref{611}) with $a_{1}=N$ and $a_{i}=0$ for $i>1$, 
we can obtain the eigenvalue and degeneracy of the system:%
\begin{equation}
E_{n}=\frac{1}{2}n^{2}-\frac{1}{2}n,
\end{equation}
\begin{equation}
\omega_{n}=\frac{n!%
{\displaystyle\prod\limits_{l=1}^{m-1}}
\left(  l+n\right)  ^{2}}{\left(  m-1\right)  !\left(  n+m-1\right)  !}.
\end{equation}

\textit{A fermionic realization: an example. }By expressing the generators of $U\left(  m\right)  $ in
terms of the Fermi operators $b_{i}^{\dagger}$ and $b_{j}$ with commutation
relation%
\begin{align}
\left\{  b_{i},b_{j}^{\dagger}\right\}   &  =\delta_{ij},\\
\left\{  b_{i}^{\dagger},b_{j}^{\dagger}\right\}   &  =\left\{  b_{i}%
,b_{j}\right\}  =0,
\end{align}
the Hamiltonian of the system, Eq. (\ref{haha}), can be written as%
\begin{equation}
H=\frac{1}{2}\sum_{i,j=1}^{m}b_{i}^{\dagger}b_{j}b_{j}^{\dagger}b_{i}-\frac
{m}{2}\sum_{i=1}^{m}b_{i}^{\dagger}b_{i}%
\end{equation}
The eigenvalue and degeneracy, by using Eqs. (\ref{610}) and (\ref{611}) with $a_{i}=a_{j}=1$, 
are expressed as%
\begin{equation}
E_{n}=-\frac{n^{2}}{2}+\frac{1}{2}n,
\end{equation}
\begin{equation}
\omega_{n}=\frac{m!}{\left(  m-n\right)  !n!}.
\end{equation}

Notice that, in Bose and Fermi cases, such system have no-fixed particle numbers.

\section{The degeneracy of eigenstates of a lattice model}
In a lattice model, usually the nearest number of particle 
is a finite numbers, such as
3, 4, 6, and so on, due to the constraint of symmetries.
Therefore, one only considers interactions between nearest particle-pairs,
second nearest particle-pairs, and so on.
In this paper, 
by converting the lattice configurations into networks, 
we can consider models with 
arbitrary number of interacting particle-pairs. 
That is, we can consider models on networks with 
different modes of links. 
In this section, we investigate 
the relationship between degeneracy of eigenstates and the number of links in networks,
by numerically calculating the Heisenberg models and random walks on network with different number of links.
It shows that models on networks with more links tend to 
have higher degeneracy of eigenstates.

\subsection{Heisenberg models and random walks on networks with different numbers of links}
Various models can be considered on a network or lattice configurations.
In this section, we consider Heisenberg models and random walks 
on networks with different modes of links.

\begin{figure}[H]
\centering
\includegraphics[width=1.0\textwidth]{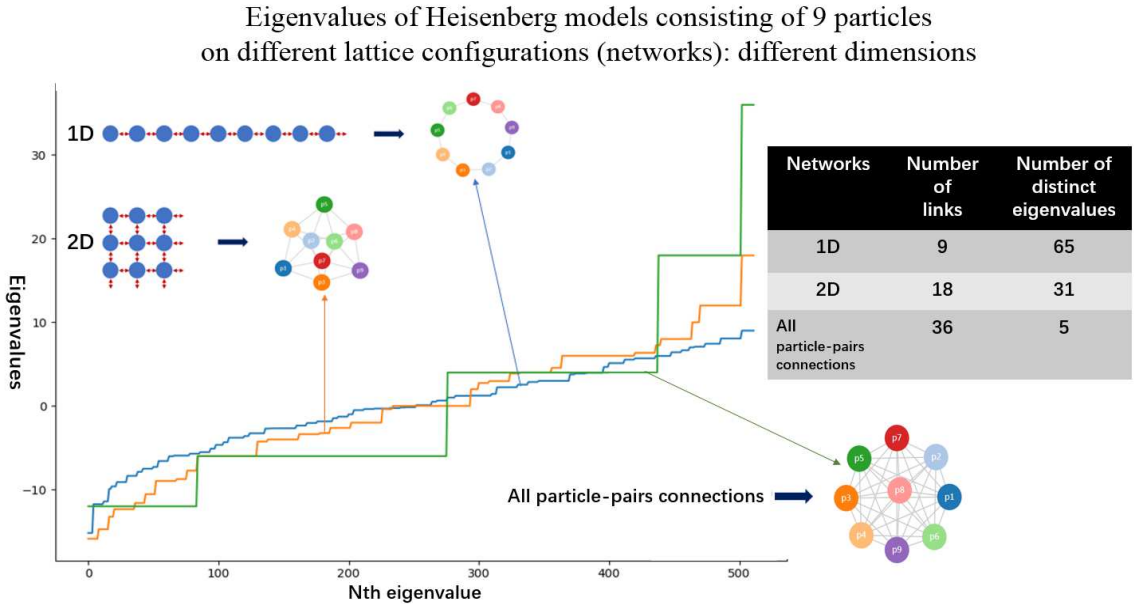}
\caption{The eigenvalue of Heisenberg model consisting of $9$ particles in different dimensions.
as shown, $1$D represents a line configuration, $2$D represents a plain configuration.}
\label{Eigenvalue_dim_dimension}
\end{figure}

\begin{figure}[H]
\centering
\includegraphics[width=1.0\textwidth]{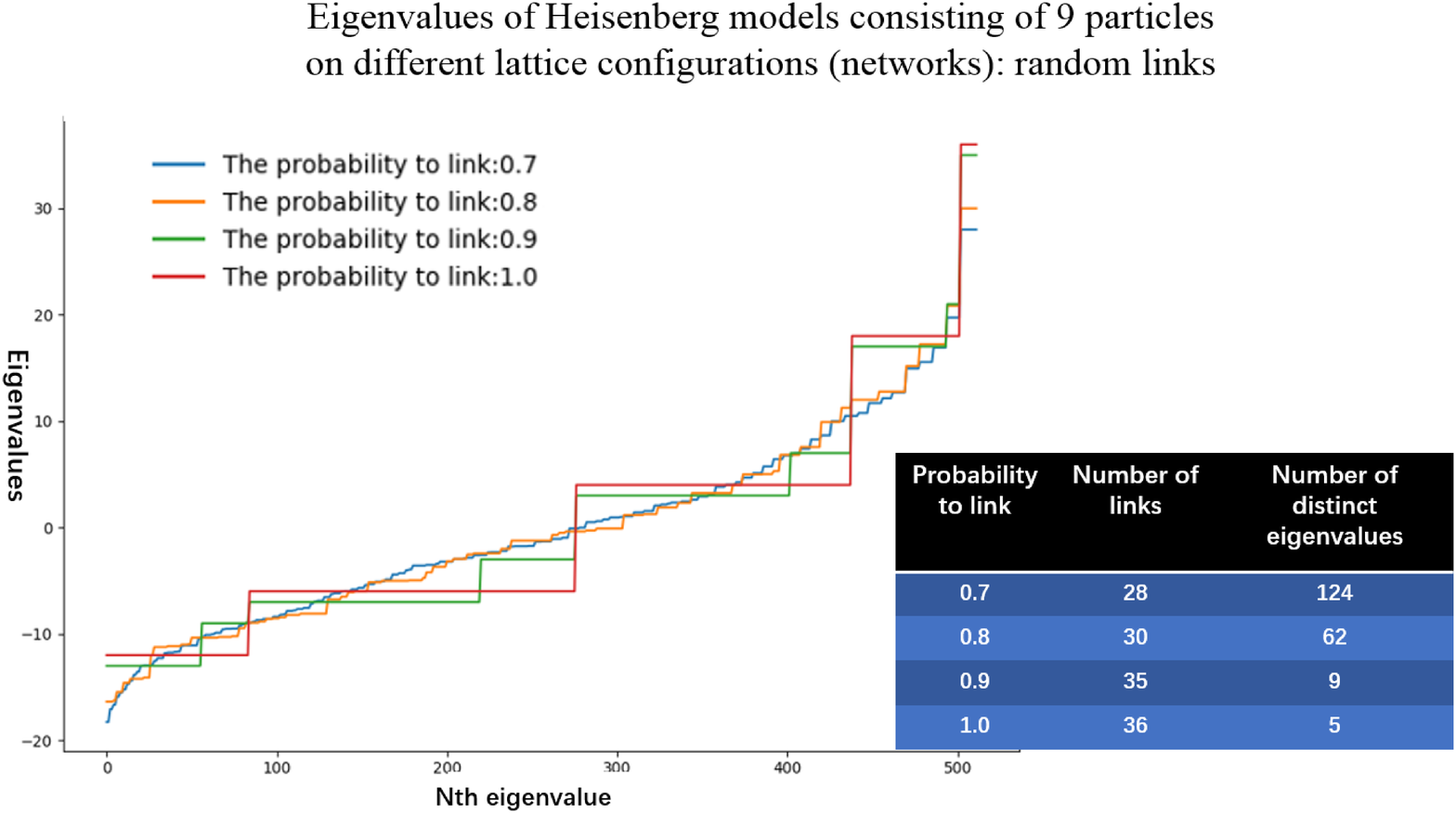}
\caption{The eigenvalue of Heisenberg model consisting of $9$ particles on networks with random links.
In a network, for any two particles, there is a probability to form a link between them.}
\label{Eigenvalue_dim_random}
\end{figure}

\textit{Heisenberg models.}
In this section, we study the isotropic Heisenberg models with
Hamiltionain $H$ reads
\begin{equation}
H=\sum_{}\frac{1}{2}\mathbf{S}_{i}\cdot\mathbf{S}_{j},\label{numer}%
\end{equation}
where $\mathbf{S}_{i}$ is the spin of $ith$ particle and $\sum_{}$ 
represents summation over particle-pairs with links. 
The eigenvalue spectrum of Heisenberg models, Eq. (\ref{numer}), in different dimensions 
and on networks with random links are
given in Figs (\ref{Eigenvalue_dim_dimension}) and (\ref{Eigenvalue_dim_random}). 
 
\begin{figure}[H]
\centering
\includegraphics[width=1.0\textwidth]{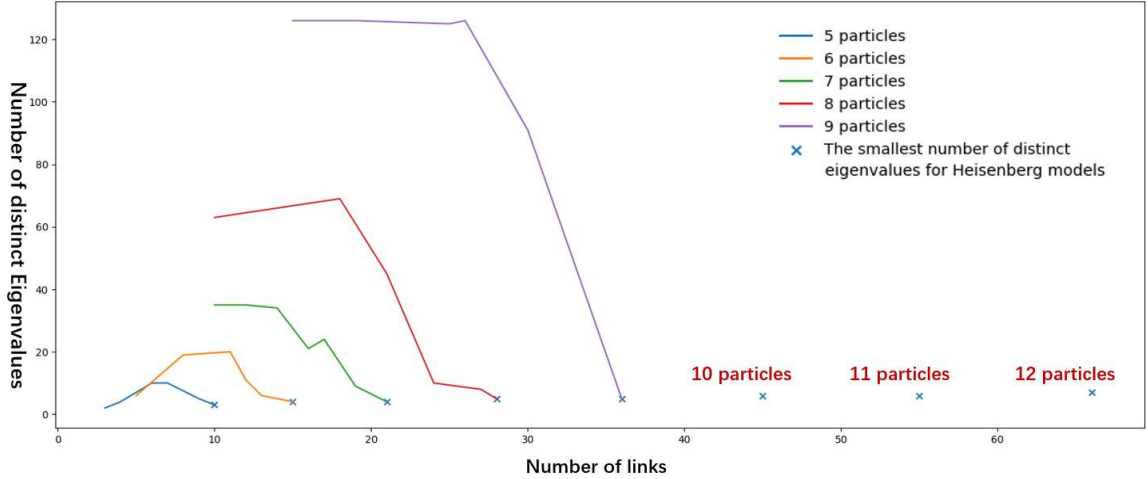}
\caption{degeneracy of eigenstates of Heisenberg models consisting of different numbers
of particles on lattice with 
 different numbers of links.
degeneracy here is revealed on the number of distinct eigenvalues.}
\label{Eigenvalue_de}
\end{figure}

Fig. (\ref{Eigenvalue_de}) shows 
the number of distinct eigenvalues of Heisenberg models 
consisting of $N$ particles
on networks with different number of links. 
It shows the tendency that degeneracy of eigenstates increases with the increase of 
the number of links when the number of links is relatively large. 

To be notice that the number of distinct eigenvalues is not strictly monotonic decreasing 
as the number of links in the networks, see Fig. (\ref{Eigenvalue_de}),
because the number of links and how the links are formed decide the eigenvalues simultaneously.
Here, we only consider the number of links in the networks.

\begin{figure}[H]
\centering
\includegraphics[width=1.0\textwidth]{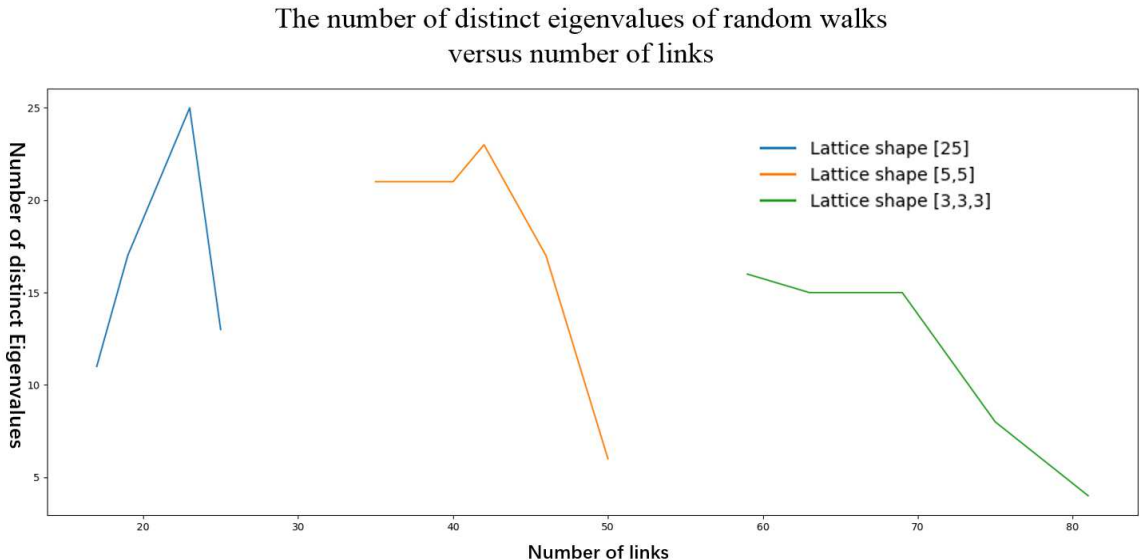}
\caption{degeneracy of eigenstates of random walks on lattice configurations with 
 different numbers of links.
degeneracy here is revealed on the number of distinct eigenvalues.}
\label{Eigenvalue_de_random}
\end{figure}

\textit{The random walk.}
Here, we construct random walks on networks with different modes of links. 
For example, the matrix of a random walk on a one-dimensional lattice configuration
consists of $5$ positions reads%
\begin{equation}
\left(
\begin{array}
[c]{ccccc}%
0 & \frac{1}{2} & 0 & 0 & \frac{1}{2}\\
\frac{1}{2} & 0 & \frac{1}{2} & 0 & 0\\
0 & \frac{1}{2} & 0 & \frac{1}{2} & 0\\
0 & 0 & \frac{1}{2} & 0 & \frac{1}{2}\\
\frac{1}{2} & 0 & 0 & \frac{1}{2} & 0
\end{array}
\right)  ,
\end{equation}
where periodic boundary conditions are applied. On a one-dimensional lattice configuration,
one can only go left or right with equal probability $1/2$.
The random walk on a one-dimensional lattice configuration with a hole at position
$1$ reads%
\begin{equation}
\left(
\begin{array}
[c]{ccccc}%
0 & 0 & 0 & 0 & 0\\
0 & 0 & 1 & 0 & 0\\
0 & \frac{1}{2} & 0 & \frac{1}{2} & 0\\
0 & 0 & \frac{1}{2} & 0 & \frac{1}{2}\\
0 & 0 & 0 & 1 & 0
\end{array}
\right)  ,
\end{equation}
where a hole means that any links to the hole is forbidden. Under such assumptions, 
one can not reach a hole.
Thus, the probability to reach a hole is $0$.
as shown in Table. (\ref{table1}) and Fig. (\ref{Eigenvalue_de_random}),
especially in three-dimension spaces, degeneracy of eigenstates
increases with the number of links in the networks. 
The number of links is adjusted by changing the number of holes in the network.

\begin{table}[H]
\label{table1}
\caption{The number of distinct eigenvalues of random walks on different lattice configurations.}
\centering
\begin{tabular}{lllll} 
\hline   
$\text{Shape}$ & $\text{Position of holes}$ & $\text{Number of links}$& $\text{Number of distinct eigenvalues}$\\  
\hline   
$1\times25$ & $-$ & $25$& $13$\\  
$1\times25$ & $4$ & $23$& $25$\\  
$1\times25$ & $2$ and $9$ & $21$& $21$\\  
$1\times25$ & $3$, $8$, and $13$ & $19$& $17$\\  
\hline 
$5\times5$ & $-$ & $50$& $6$\\  
$5\times5$ & $(1,4)$ & $46$& $17$\\  
$5\times5$ & $(1,2)$ and $(2,4)$ & $42$& $23$\\  
$5\times5$ & $(1,3)$, $(2,3)$, and $(3,3)$ & $40$& $21$\\  
\hline 
$3\times3\times3$ & $-$ & $81$& $4$\\  
$3\times3\times3$ & $(2,1,1)$ & $75$& $8$\\  
$3\times3\times3$ & $(1,2,1)$ and $(3,3,1)$ & $69$& $15$\\  
$3\times3\times3$ & $(2,2,1)$, $(3,3,1)$, $(3,3,2)$, and $(2,3,3)$ & $59$& $16$\\ 
\hline  
\end{tabular}
\end{table}

\subsection{The highest degeneracy of eigenstates of a lattice model}
In Sec. 4.1, it shows that degeneracy of
eigenvalues increases with the number 
of links in a network. That is, models on configurations 
with larger number of interacting particle-pairs 
have higher degeneracy of eigenstates. 
In this section, we propose the assumption that 
the lattice model with interactions between all particle-pairs, such as the generalized model proposed in Sec. 3.2,
should have the highest degeneracy of eigenstates.

\begin{figure}[H]
\centering
\includegraphics[width=1.0\textwidth]{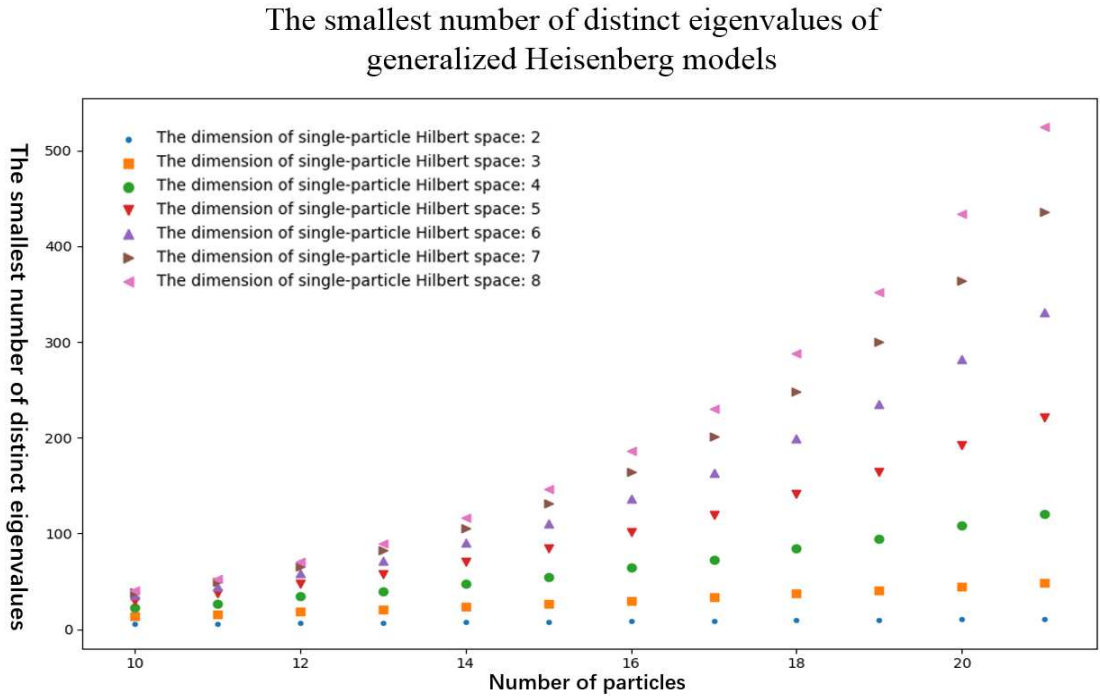}
\caption{The smallest number of distinct eigenvalues of the generalized Heisenberg model.}
\label{smallest}
\end{figure}

\textit{The highest degeneracy of eigenstates of the generalized Heisenberg model.}
For the generalized Heisenberg model 
consisting of $N$ particles with the dimension of 
single-particle-Hilbert space $m$, by using Eq. (\ref{610}), 
the number of distinct eigenvalues
is $P(N,m)$, where $P(N,m)$ is the restrict integer partition number \cite{zhou2018statistical}
that is the number of ways to express $N$
as sum of other integers with the number of summands no larger than $m$. 
It is because one irreducible representation gives a distinct 
eigenvalue of Casimir operator and thus gives a distinct eigenvalue of the system.
The irreducible representation is labeled by a set of number 
$\left(  a\right)  =\left(  a_{1}%
,a_{2},...,a_{m}\right)  $ with $a_{1}\geqslant a_{2}\geqslant...\geqslant
a_{m}\geqslant0$ \cite{Iachello2006Lie}. 
The summation of $\left(  a\right)  $ is $N$, thus, the number of $\left(  a\right)$
equals the number of ways to represent $N$ in terms of other integers the number of 
summands smaller than $m$, that is $P(N,m)$. 
as shown in Fig. (\ref{smallest}). 

For these cases, the average degeneracy of an energy 
level reads $m^{N}/P(N,m)$. For example, setting $m=3$ and $N=6$, the smallest number 
of distinct eigenvalues is $P(N,m)=7$, the average degeneracy is $104.14$. as shown in Table. (\ref{av}),
the average degeneracy of an energy level in creases fast with the number 
of particles or the dimension of single-particle Hilbert space.

\begin{table}[H]
\label{av}
\caption{The average degeneracy of an energy level at the upper limit of energy level degeneracy.}
\centering
\begin{tabular}{lllll} 
\hline   
$\text{The dimension of single-particle Hilbert space}$ & $\text{Number of particles}$ & $\text{The average degeneracy}$\\  
\hline   
$2$ & $5$ & $10.7$\\  
$2$ & $6$ & $16.0$\\ 
$2$ & $7$ & $32.0$\\ 
$2$ & $8$ & $51.2$\\  
$2$ & $9$ & $102.4$\\  
$2$ & $10$ & $170.7$\\  
$3$ & $5$ & $48.6$\\   
$3$ & $6$ & $104.1$\\  
$3$ & $7$ & $273.4$\\  
$3$ & $8$ & $656.1$\\ 
$3$ & $9$ & $1640.3$\\  
$3$ & $10$ & $4217.8$\\
\hline  
\end{tabular}
\end{table}

\section{Conclusions and discussions}
The difficulty of solving the interacting quantum many-body system 
is due to the properties of the space configuration \cite{wen2004quantum,flaschner2018observation,stenzel2019quantum}
and the interaction mode between inter-particles including the classical interaction 
and the quantum exchange interaction \cite{dai2007interacting,dai2005hard,zhou2018canonical,zhou2018calculating}. 
Few models of quantum interacting 
many-body systems could be solved exactly in reality.
Common practices are simplifying the models or using approximation methods
instead of solving the system directly, we conclude that to investigate
the system from the perspective of properties of the space
and symmetries of the system is useful and effective.

In this paper, we consider models on lattice configurations
with arbitrary numbers of interacting particle-pairs
by converting the lattice configurations
into networks with different modes of links. 
There are three main purposes of our work:
(1) we introduce a group theory method to solve the Heisenberg model on lattice configurations 
with interactions 
between all particle-pairs
by revealing the relation between the Casimir operator of the unitary group and the
conjugacy-class operator of the permutation group.
(2) We generalize the Heisenberg model by this proposed relation and thus 
give a series of exactly solvable models.
(3) We show that degeneracy of the
eigenvalue increases with the number of interacting particle-pairs.
Thus, there is a highest degeneracy of eigenstates of a 
lattice model. 
The generalized Heisenberg model is a model on lattice configurations 
with interactions between
all particle-pairs, and thus have the highest degeneracy of the
eigenvalue among lattice models.
The smallest number of distinct eigenvalues for 
the generalized Heisenberg model
is a restrict integer partition function.

\section{Acknowledgments}
We are very indebted to Dr Yong Xie  and Dr Guanwen Fang for their encouragements.
This work is supported by the Fundamental Research Funds For the Central Universities No.2020JKF306.












\end{document}